\documentclass[apex,twocolumn]{jjap3_published}

\usepackage{newtxtext,newtxmath}
\usepackage{bm}

\title{%
Theoretical evaluation of the screening-current-induced magnetic field 
in superconducting coils with tape wires
}
\author{%
Yasunori Mawatari
}
\inst{%
National Institute of Advanced Industrial Science and Technology (AIST), 
Tsukuba, Ibaraki 305-8568, Japan
}

\abst{%
We theoretically investigate the physical mechanism of the screening-current-induced field (SCIF) in solenoid coils wound with superconducting tape wires. 
We derive the direct relationship between the SCIF and the magnetization of tape wires, and a scaling law for the SCIF and the coil dimensions is demonstrated. 
A simple analytical expression of the SCIF is obtained as functions of current load factor, tape wire width, and the coil dimensions. 
We verify that the published data for the precise numerical computation of SCIF are roughly fitted by our theoretical results for flat coils where the height is smaller than the outer diameter. 
}

\begin{document}
\maketitle

High-temperature superconducting (HTS) coils have been developed for high-field magnet applications,~\cite{Trociewitz_11,Maeda_13} such as magnetic resonance imaging (MRI),~\cite{Miyazaki_16,Yokoyama_17} nuclear magnetic resonance (NMR),~\cite{Yanagisawa_14,Nishijima_16} and accelerator magnets.~\cite{Amemiya_12,Ueda_13,Rossi_15}  
Rare-earth HTS tape wires have excellent superconducting properties at temperatures that can be accessed by using cryocoolers without liquid helium, and thus these tape wires are promising candidates for high-field magnets.~\cite{Trociewitz_11,Maeda_13,Miyazaki_16,Yokoyama_17,Yanagisawa_14,Benkel_17}
However, in tape wires, large magnetization arises from the screening currents in the wide surfaces, resulting in fatal irregular magnetic fields (i.e., spatially nonuniform and temporally unstable magnetic fields).~\cite{Hemmi_07,Amemiya_08,Ahn_09,Koyama_09} 
A few techniques have been proposed to suppress the effects of the screening-current-induced field (SCIF).~\cite{Yanagisawa_09,Kajikawa_11,Yanagisawa_12,Ueda_15,Yanagisawa_15}
Numerical computation is used for designing HTS magnets, but huge computational resources and highly advanced numerical techniques are required for precise evaluation of the SCIF~\cite{Amemiya_08,Ueda_15,Pardo_16}. 
Theoretical investigation is thus desired to unveil the underlying physics of the SCIF and to provide a simple method to evaluate the SCIF. 

In this paper, we theoretically investigate the SCIF in solenoid coils with superconducting tape wires, and we propose analytical expressions for evaluation of the SCIF. 
The simple analytical results allows us to evaluate the SCIF roughly but quite easily, and are useful for designing HTS coils and magnets.

We consider the general expression for the magnetic field in superconducting coils including the effect of the SCIF. 
The vector potential, $\bm A$, due to the current density, $\bm J$, flowing in wires in coils is generally given by ${\bm A}({\bm r})=(\mu_0/4\pi)\int d V'{\bm J}({\bm r}')/|{\bm r}-{\bm r'}|$, where $\mu_0$ is the vacuum permeability, and the volume integral, $\int dV'$, is calculated over the conductor area in a coil. 
When we regard the discrete windings of tape wires as a continuum (i.e., a homogeneous medium), the current density may be simply averaged out as the mean transport current density, ${\bm J}\simeq {\bm J}_{\rm t}$. 
In this naive approximation, however, the SCIF effect is neglected. 
The SCIF effect can be taken into account (even in the continuum model) in terms of the magnetization current, as in magnetic materials~\cite{Jackson_75}. 
The magnetization current density, ${\bm J}_{\rm m}$, has a volume contribution, $\nabla\times{\bm M}$, and a surface contribution, ${\bm M}\times{\hat{\bm n}}$, where $\bm M$ is the magnetization and ${\hat{\bm n}}$ is the normal unit vector. 
The vector potential due to the transport current density, ${\bm J}_{\rm t}$, and the magnetization current density, ${\bm J}_{\rm m}$, is then given by 
\begin{equation}
	{\bm A}({\bm r}) \simeq \frac{\mu_0}{4\pi}\int 
		\frac{{\bm J}_{\rm t}({\bm r}')+{\bm J}_{\rm m}({\bm r}')}{|{\bm r}-{\bm r}'|}d V'
	= {\bm A}_{\rm t} +{\bm A}_{\rm m}, 
\label{eq:A=Jt+Jm}
\end{equation}
where ${\bm A}_{\rm t}$ and ${\bm A}_{\rm m}$ are the contributions from ${\bm J}_{\rm t}$ and ${\bm J}_{\rm m}$, respectively, 
\begin{align}
	{\bm A}_{\rm t}({\bm r}) &= \frac{\mu_0}{4\pi}\int 
		\frac{{\bm J}_{\rm t}({\bm r}')}{|{\bm r}-{\bm r}'|}d V' , 
\label{eq:At}\\
	{\bm A}_{\rm m}({\bm r}) &= \frac{\mu_0}{4\pi}\int 
		\frac{{\bm M}({\bm r}')\times ({\bm r}-{\bm r}')}{%
		|{\bm r}-{\bm r}'|^3}d V' . 
\label{eq:Am}
\end{align}
The resulting magnetic field, ${\bm B}=\nabla\times{\bm A}$, in coils is given as the sum of the contribution from the transport current, ${\bm B}_{\rm TC}=\nabla\times{\bm A}_{\rm t}$, and that from the screening current in wires (i.e., magnetization current), ${\bm B}_{\rm SC}=\nabla\times{\bm A}_{\rm m}$. 
Note that ${\bm B}_{\rm SC}$ corresponds to the SCIF.

In the cylindrical coordinates, $(r,\theta,z)$, we consider the axisymmetric vector potential, ${\bm A}= A_{\theta,k}(r,z)\hat{\bm\theta}$, due to the azimuthal current density, ${\bm J}= J_{\theta,k}(r,z)\hat{\bm\theta}$, flowing in a single-turn coil of a tape wire. 
The tape width is $w$, the superconducting-layer thickness is $d_{\rm s}$, the coil radius is $r_k$, and the center of the coil is at $z=z_k$. 
The index $k$ is used to calculate the magnetic field in a multi-turn solenoid coil. 
Near the central $z$ axis, $r\ll r_k$, the vector potential, $A_{\theta,k}$, is given by 
\begin{align}
	A_{\theta,k}(r,z) &\simeq \mu_0 d_{\rm s} \int_{z_k-w/2}^{z_k+w/2} 
		\frac{r r_k^2 J_{\theta,k}(r_k,z')}{%
		4\left[ r_k^2+(z-z')^2 \right]^{3/2}} d z'  \nonumber\\
	&\simeq \frac{r r_k^2}{4\left[ r_k^2+(z-z_k)^2 \right]^{3/2}} 
		\mu_0 I_{\rm t} 
\nonumber\\
	&\quad	-\frac{3r r_k^2(z-z_k)}{4\left[ r_k^2+(z-z_k)^2 \right]^{5/2}} 
		\mu_0 M_r(r_k,z_k)wd_{\rm s} + \cdots , 
\label{eq:Ak_single}
\end{align}
where $A_{\theta,k}$ is expressed as the multipole expansion. 
In the last expression of Eq.~\eqref{eq:Ak_single}, the first term corresponds to ${\bm A}_{\rm t}$ given by Eq.~\eqref{eq:At} and the second term corresponds to ${\bm A}_{\rm m}$ given by Eq.~\eqref{eq:Am}. 
The transport current, $I_{\rm t}$, the mean transport current density, $J_{\rm t}$, and the radial magnetization, $M_r$, are defined by 
\begin{align}
	I_{\rm t} = J_{\rm t} wd_{\rm s} 
	&= d_{\rm s} \int_{z_k-w/2}^{z_k+w/2} J_{\theta,k}(r_k,z') d z', 
\label{eq:It_definition}\\
	M_r(r_k,z_k) 
	&= \frac{1}{w}\int_{z_k-w/2}^{z_k+w/2} J_{\theta,k}(r_k,z')(z_k-z') d z' . 
\label{eq:Mr_definition}
\end{align}
For HTS tape wires of $d_{\rm s}\ll w$, the thicknesswise distribution of $J_{\theta,k}$ is assumed to be uniform and we consider the widthwise distribution in response to the radial component of the magnetic field.
Based on Bean's model with constant critical current density $J_{\rm c}$~\cite{Bean_62,Bean_64}, the magnetization, $M_r$, has a maximum, $|M_r| \leq J_{\rm c} w/4$ for $I_{\rm t}=0$. 
For $I_{\rm t}\neq 0$, on the other hand, the maximum magnetization is given by 
\begin{equation}
	|M_r| \leq M_{\rm max}\equiv \frac{J_{\rm c}w}{4}
		\left[ 1- \left(\frac{J_{\rm t}}{J_{\rm c}}\right)^2 \right] . 
\label{eq:M_max}
\end{equation}
$M_r= \mp M_{\rm max}$ is obtained from Eq.~\eqref{eq:Mr_definition} with the following $J_{\theta,k}$ distribution: $J_{\theta,k}= \pm J_{\rm c}$ for $-w/2 < z'-z_k < -(w/2)J_{\rm t}/J_{\rm c}$ and $J_{\theta,k}= \mp J_{\rm c}$ for $-(w/2)J_{\rm t}/J_{\rm c}<z'-z_k<+w/2$. 
The axial magnetic field, $B_{z,k}=\partial(rA_{\theta, k})/\partial r/r$, at the origin, $(r,z)=(0,0)$, due to the $k$th single-turn coil is then given by 
\begin{align} 
	B_{z,k}(0,0) &\simeq \frac{r_k^2}{2(r_k^2+z_k^2)^{3/2}} 
		\mu_0 I_{\rm t} 
\nonumber\\
	&\quad -\frac{3r_k^2 z_k}{2(r_k^2+z_k^2)^{5/2}} 
		\mu_0 M_r(r_k,z_k) wd_{\rm s} + \cdots , 
\label{eq:Bk_single}
\end{align}
where the second term of the right-hand side of Eq.~\eqref{eq:Bk_single} proportional to $M_r$ corresponds to the SCIF~\cite{Kajikawa_15}.

We consider the magnetic field in a multi-turn solenoid coil of inner radius $a_1$, outer radius $a_2$, and height $2b$. 
The multiple layers of tape wires are regarded as a set of concentric circular coils, and the vector potential for a multi-turn solenoid coil can be calculated as the sum of the contributions from the $k$th single-turn coil, 
\begin{align}
	A_{\theta}(r,z) &= \sum_k A_{\theta,k}(r,z)
\nonumber\\
	&\simeq \int_{a_1}^{a_2}d r_k \int_{-b}^{+b}d z_k 
	\frac{\lambda_{\rm s}}{wd_{\rm s}} \ A_{\theta,k}(r,z), 
\label{eq:A_multi-turn}
\end{align}
where the sum with respect to the layer index, $k$, of the single-turn coils are approximated as the integral with respect to $r_k$ and $z_k$. 
In Eq.~\eqref{eq:A_multi-turn}, $\lambda_{\rm s} =Nwd_{\rm s}/2(a_2-a_1)b$ is the ratio of the total superconductor area to the cross-sectional area of a coil $a_1<r<a_2$ and $|z|<b$, where $N$ is the total winding number. 
The factor, $\lambda_{\rm s}/wd_{\rm s} =N/2(a_2-a_1)b$, corresponds to the winding density of tape wires (i.e., winding number per unit area).
Here, we focus on the magnetic field at the center of a coil, which is derived from Eqs.~\eqref{eq:Ak_single} and \eqref{eq:A_multi-turn}, 
\begin{equation}
	B_z(0,0) = B_{\rm TC} + B_{\rm SC}, 
\label{eq:coil-center-field}
\end{equation}
where $B_{\rm TC}$ is the transport-current-induced field and $B_{\rm SC}$ is the SCIF, 
\begin{align}
	B_{\rm TC} &\simeq \mu_0 
	\int_{a_1}^{a_2}d r_k \int_{-b}^{+b}d z_k 
	\frac{r_k^2}{2(r_k^2+z_k^2)^{3/2}} \lambda_{\rm s} J_{\rm t} ,  
\label{eq:Btc}\\
	B_{\rm SC} &\simeq \mu_0 
	\int_{a_1}^{a_2}d r_k \int_{-b}^{+b}d z_k 
	\frac{3r_k^2 z_k}{2(r_k^2+z_k^2)^{5/2}} \lambda_{\rm s} M_r(r_k,z_k) . 
\label{eq:Bsc}
\end{align}
For uniformly wound coils (i.e., constant $\lambda_{\rm s}$), $B_{\rm TC}$ is proportional to $\lambda_{\rm s} J_{\rm t}$, 
\begin{equation}
	B_{\rm TC} \simeq \mu_0\lambda_{\rm s} J_{\rm t} a_1 F_{\rm TC} , 
\label{eq:Btc-Ftc}
\end{equation}
where the coil geometry factor, $F_{\rm TC}$, is determined by the coil dimensions~\cite{Wilson83}, 
\begin{align}
	F_{\rm TC}(a_1,a_2,b) &= \frac{1}{a_1} 
	\int_{a_1}^{a_2}d r_k \int_{-b}^{+b}d z_k 
	\frac{r_k^2}{2(r_k^2+z_k^2)^{3/2}} 
\nonumber\\
	&= \frac{b}{a_1} \ln\left( 
	\frac{a_2+\sqrt{a_2^2+b^2}}{a_1+\sqrt{a_1^2+b^2}} \right) . 
\label{eq:Ftc}
\end{align}

We discuss the scaling law for the SCIF and the coil dimensions based on Eqs.~\eqref{eq:Btc}--\eqref{eq:Ftc}. 
We consider the case where the coil dimensions are enlarged similarly by a factor of $f$, 
$(a_1,a_2,b)\to (fa_1,fa_2,fb)$, 
whereas the tape width, $w$, the ratio of the total superconductor area to the conductor area, $\lambda_{\rm s}$, and the transport current, $I_{\rm t}$, are fixed. 
In this case, the coil geometry factor given by Eq.~\eqref{eq:Ftc} is invariant, and the transport-current-induced magnetic field given by Eq.~\eqref{eq:Btc-Ftc} changes as $B_{\rm TC}\to fB_{\rm TC}$, because $a_1$ appears in Eq.~\eqref{eq:Btc-Ftc}. 
In other words, the transport-current-induced field is proportional to the coil size. 
The tape wires are exposed to magnetic fields that are larger by a factor of $f$, but the magnetization, $M_r$, may be regarded as roughly invariant because of the nonlinear magnetic response of superconductors. 
With this assumption [i.e., $M_r(r_k,z_k)\sim M_r(fr_k,fz_k)$] and the variable transformation, $(r_k,z_k)\to (fr_k,fz_k)$ in Eq.~\eqref{eq:Bsc}, the SCIF given by Eq.~\eqref{eq:Bsc} is expected to be almost invariant, $B_{\rm SC}\to B_{\rm SC}$. 
The rough invariance of $B_{\rm SC}$ as the coil dimensions change has been numerically verified in Fig.~2(a) of Ref.~\citen{Yanagisawa_11}.
The resulting ratio of $B_{\rm SC}$ to $B_{\rm TC}$, called the SCIF ratio, is roughly proportional to $1/f$ as $|B_{\rm SC}/B_{\rm TC}| \to (1/f)|B_{\rm SC}/B_{\rm TC}|$.

The scaling law discussed above can also be derived simply from the dimensional analysis of Amp\`ere's law, $\nabla\times\bm{B} =\mu_0\bm{J}$, as follows. 
The magnetic induction, $\bm{B}= \bm{B}_{\rm TC} +\bm{B}_{\rm SC}$, is composed of the transport-current induced field, $\bm{B}_{\rm TC}$, and the SCIF, $\bm{B}_{\rm SC}$. The current density, $\bm{J}= \bm{J}_{\rm TC} +\bm{J}_{\rm SC}$, is composed of the transport current density, $\bm{J}_{\rm TC}$, and the screening current density, $\bm{J}_{\rm SC}= \nabla\times\bm{M}$, where the screening current density is regarded as the magnetization current density due to the magnetization, $\bm{M}$. 
Amp\`ere's law is then decomposed as $\nabla\times\bm{B}_{\rm TC} =\mu_0\bm{J}_{\rm TC}$ and $\nabla\times\bm{B}_{\rm SC} =\mu_0\nabla\times\bm{M}$. 
For the dimensional analysis, the nabla operator can be regarded as having a dimension of $|\nabla|\sim 1/{\cal L}$, where ${\cal L}$ represents the length dimension. 
Amp\`ere's law for the transport current, $\nabla\times\bm{B}_{\rm TC} =\mu_0\bm{J}_{\rm TC}$, leads to the dimensional relationship, $|\bm{B}_{\rm TC}|/{\cal L} \sim\mu_0|\bm{J}_{\rm TC}|$. 
Because $\bm{J}_{\rm TC}$ is independent of ${\cal L}$, we have $|\bm{B}_{\rm TC}| \propto {\cal L}$. 
Amp\`ere's law for the screening current, $\nabla\times\bm{B}_{\rm SC} =\mu_0\nabla\times\bm{M}$, leads to the dimensional relationship, $|\bm{B}_{\rm SC}|/{\cal L} \sim\mu_0|\bm{M}|/{\cal L}$. 
Because $|\bm{M}|$ is roughly independent of ${\cal L}$, the SCIF, $|\bm{B}_{\rm SC}| \sim\mu_0|\bm{M}|$, is also roughly independent of ${\cal L}$. 
Thus, we arrive at the scaling relations for ${\cal L}\to f{\cal L}$, $|\bm{B}_{\rm TC}|\to f|\bm{B}_{\rm TC}|$ and $|\bm{B}_{\rm SC}|\to |\bm{B}_{\rm SC}|$.

We derive a simple expression of the SCIF. 
The distribution of $M_r(r_k,z_k)$ in Eq.~\eqref{eq:Bsc} is complicated and its accurate evaluation requires an advanced, powerful numerical technique~\cite{Amemiya_08,Ueda_15,Pardo_16}. 
We evaluate $B_{\rm SC}$ roughly by using Bean's critical state model with constant critical current density $J_{\rm c}$~\cite{Bean_62,Bean_64} and assuming the following simple distribution of $M_r$. 
We consider the case where the azimuthal transport current, $I_{\rm t}$, is increased monotonically, and the radial component of the magnetic field, $B_r$, arises in a coil, inducing the screening current and the radial magnetization, $M_r$, of the tape wires. 
In the upper half of a coil, $z>0$, the radial field is positive, $B_r>0$, and the radial magnetization is negative, $M_r<0$. 
In the lower half of a coil, $z<0$, $B_r<0$, and $M_r>0$. 
Although the accurate behavior of $M_r$ is complicated, the maximum of $|M_r|$ is simply given by Eq.~\eqref{eq:M_max}. 
We assume that all windings of the tape wires are fully magnetized responding to $B_r$, and we use the simple distribution of the magnetization for ascending $I_{\rm t}$, 
\begin{equation}
	M_r \sim \left\{%
	\begin{aligned}
		&-M_{\rm max} & \mbox{for } z_k >0 , \\
		&+M_{\rm max} & \mbox{for } z_k <0 . 
	\end{aligned}\right.
\label{eq:Mr=Mmax}
\end{equation}
This fully magnetized wire model may be acceptable for flat coils of $b/a_2 <1$, whereas the model overestimates the magnetization for tall coils of $b/a_2 > 1$, because $|M_r|\ll M_{\rm max}$ for $|z_k|\ll b$.
The maximum magnetization given by Eq.~\eqref{eq:M_max} contains the local critical current density $J_{\rm c}$ in tape wires, and is rewritten as 
\begin{equation}
	M_{\rm max}= \frac{J_{\rm t}w}{4}
		\left( \frac{I_{\rm c,tape}}{I_{\rm t}} 
		- \frac{I_{\rm t}}{I_{\rm c,tape}} \right) . 
\label{eq:Mmax-Ic,tape}
\end{equation}
The critical current of tape wires, $I_{\rm c,tape}= J_{\rm c}wd_{\rm s}$, is spatially distributed along the wire length, and $I_{\rm c,tape}$ depends on the position $(r_k,z_k)$ where tape wires are situated in a coil. 
The spatial distribution of $I_{\rm c,tape}$ is due to the inhomogeneity of the superconducting characteristics and to the dependence on the local magnetic field produced in a coil. 
Therefore, the critical current of a coil (i.e., the critical current of the total wires), $I_{\rm c,coil}$, is determined by the minimum of the local $I_{\rm c,tape}$. 
Using the parameter $\alpha_{\rm c} =I_{\rm c,tape}/I_{\rm c,coil} >1$, we further rewrite Eq.~\eqref{eq:Mmax-Ic,tape} as 
\begin{equation}
	M_{\rm max}= \frac{J_{\rm t}w}{4}
		\left( \frac{\alpha_{\rm c} I_{\rm c,coil}}{I_{\rm t}} 
		- \frac{I_{\rm t}}{\alpha_{\rm c} I_{\rm c,coil}} \right) . 
\label{eq:Mmax-Ic,coil}
\end{equation}
Although the parameter $\alpha_{\rm c}\propto I_{\rm c,tape}$ is spatially distributed in a coil, we regard $\alpha_{\rm c}$ to be a spatially averaged value of $I_{\rm c,tape}/I_{\rm c,coil}$ in a coil for simplicity. 
Figure 8(a) in Ref. \citen{Benkel_17} shows the distribution of $I_{\rm c,tape}$ from the numerical simulation taking account of the dependence of $J_{\rm c}$ upon the magnetic-field strength and angle. 
From this $I_{\rm c,tape}$ distribution, the averaged $\alpha_{\rm c}= I_{\rm c,tape}/I_{\rm c,coil}$ is estimated as $\alpha_{\rm c}\sim 1.5$. 
Although $\alpha_{\rm c}$ may depend on the dimensions of the coils and on the behavior of $J_{\rm c}$, the value of Eq.~\eqref{eq:Mmax-Ic,coil} is almost unchanged for $1<\alpha_{\rm c}<2$ except for $I_{\rm t}\sim I_{\rm c, coil}$.
With the fully magnetized wire model of Eq.~\eqref{eq:Mr=Mmax} and the spatially averaged $\alpha_{\rm c}$ of Eq.~\eqref{eq:Mmax-Ic,coil}, the SCIF given by Eq.~\eqref{eq:Bsc} is calculated as 
\begin{equation}
	B_{\rm SC} \sim -\mu_0 \lambda_{\rm s} M_{\rm max} F_{\rm SC} 
	\quad\mbox{for ascending } I_{\rm t} , 
\label{eq:Bsc=-MmaxFsc}
\end{equation}
where the coil geometry factor, $F_{\rm SC}$, is given by 
\begin{align}
	F_{\rm SC}(a_1,a_2,b) 
	&= \int_{a_1}^{a_2}d r_k \int_{-b}^{+b}d z_k 
	\frac{3r_k^2 |z_k|}{2(r_k^2+z_k^2)^{5/2}}  
\nonumber\\
	&= \frac{a_2}{\sqrt{a_2^2+b^2}} -\frac{a_1}{\sqrt{a_1^2+b^2}} 
\nonumber\\
	&\quad +\ln\left[ \frac{a_2\left(a_1+\sqrt{a_1^2+b^2}\right)}{%
	a_1\left(a_2+\sqrt{a_2^2+b^2}\right) }\right] . 
\label{eq:Fsc}
\end{align}
Equation.~\eqref{eq:Bsc=-MmaxFsc} is valid for monotonically ascending $I_{\rm t}$, whereas the $\pm$ signs of $M_r$ in Eq.~\eqref{eq:Mr=Mmax} and of $B_{\rm SC}$ in Eq.~\eqref{eq:Bsc=-MmaxFsc} are reversed for monotonically descending $I_{\rm t}$, 
\begin{equation}
	B_{\rm SC} \sim +\mu_0 \lambda_{\rm s} M_{\rm max} F_{\rm SC} 
	\quad\mbox{for descending }I_{\rm t} . 
\label{eq:Bsc=+MmaxFsc}
\end{equation}
From Eqs.~\eqref{eq:Btc-Ftc}, \eqref{eq:Mmax-Ic,coil}, \eqref{eq:Bsc=-MmaxFsc}, and \eqref{eq:Bsc=+MmaxFsc}, the SCIF ratio is given by 
\begin{equation}
	\left| \frac{B_{\rm SC}}{B_{\rm TC}} \right| 
	\sim \frac{M_{\rm max} F_{\rm SC}}{J_{\rm t}a_1 F_{\rm TC}} 
	= \left( \frac{\alpha_{\rm c} I_{\rm c,coil}}{I_{\rm t}} 
		-\frac{I_{\rm t}}{\alpha_{\rm c} I_{\rm c,coil}}\right) 
		\frac{w F_{\rm SC}}{4a_1 F_{\rm TC}} . 
\label{eq:Bsc/Btc_alpha}
\end{equation}
The ratio of the coil geometry factors, $F_{\rm SC}/F_{\rm TC}$, is determined only by the dimensions of a solenoid coil, $a_1$, $a_2$, and $b$. 
For example, for a thin, flat solenoid coil (i.e., $a_2-a_1 \ll a_1$ and $b\ll a_1$), we have $F_{\rm SC}/F_{\rm TC} \sim 3b/2a_1$.

We consider the dependence of the SCIF ratio, $|B_{\rm SC}/B_{\rm TC}| \propto w$, on the tape width, $w$, and our theoretical results are compared with the numerical results from Ueda {\em et al}.~\cite{Ueda_15}. 
\begin{figure}[tb]
 \center\includegraphics{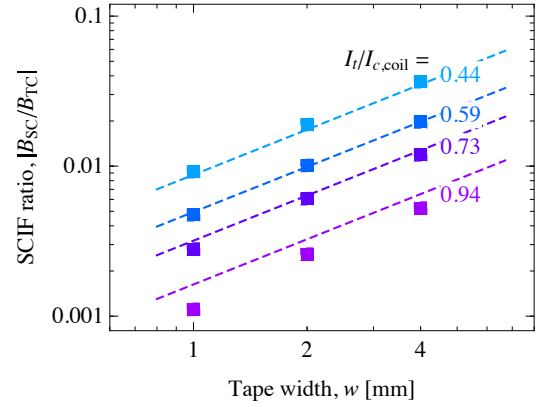}
\caption{%
SCIF ratio $|B_{\rm SC}/B_{\rm TC}|$ as a function of tape width $w$ for $I_{\rm t}/I_{\rm c,coil}= 0.44$, $0.59$, $0.73$, and $0.94$. 
The symbols show the numerical results from Ueda {\em et al}.~\cite{Ueda_15}, and the dashed lines show our theoretical results given by Eq.~\eqref{eq:Bsc/Btc_alpha} with $\alpha_{\rm c}=1.5$. 
The inner diameter, outer diameter, and height of coils are $2a_1=100\,$mm, $2a_2=170.5\,$mm, and $2b=39$--$79\,$mm, respectively~\cite{Ueda_15}. 
}\label{fig:Bsc_w} 
\end{figure}
Figure~\ref{fig:Bsc_w} shows the plot of the SCIF ratio versus tape width $w$ for various current load factors, $I_{\rm t}/I_{\rm c,coil}$. 
The numerical data (symbols) taken from Ref.~\citen{Ueda_15} clearly demonstrate that $|B_{\rm SC}/B_{\rm TC}|$ is proportional to $w$, and the data are roughly fitted by our theoretical results (dashed lines) obtained from Eq.~\eqref{eq:Bsc/Btc_alpha}.

We consider here the dependence of the SCIF ratio, $|B_{\rm SC}/B_{\rm TC}|\propto F_{\rm SC}/a_1F_{\rm TC}$, on the coil dimensions, and our theoretical results are compared with the numerical results from Yanagisawa {\em et al}.~\cite{Yanagisawa_11}.
\begin{figure}[tb]
 \center\includegraphics{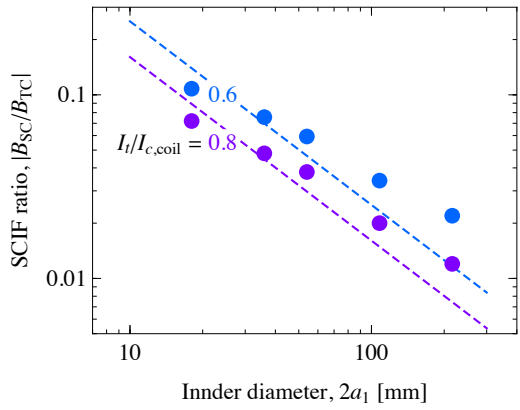}
\caption{%
SCIF ratio $|B_{\rm SC}/B_{\rm TC}|$ as a function of inner diameter $2a_1$ of coils for $I_{\rm t}/I_{\rm c,coil}=0.6$ and $0.8$. 
The symbols show the numerical results from Yanagisawa {\em et al}.~\cite{Yanagisawa_11}, and the dashed lines show our theoretical results given by Eq.~\eqref{eq:Bsc/Btc_alpha} with $\alpha_{\rm c}=1.5$. 
The tape width is $w=4\,$mm and the coil factor is $F_{\rm SC}/F_{\rm TC}=0.60$. 
}\label{fig:Bsc_a1}
\end{figure}
They numerically investigated the SCIF for five coils with similar configurations and various dimensional scaling factors, $f$. 
The inner diameter, outer diameter, and height of those coils are $(2a_1,2a_2,2b)= f\times(18\,{\rm mm},40\,{\rm mm},25\,{\rm mm})$, where $f=1$, 2, 3, 6, and 12. 
The coil geometry factors given by Eqs.~\eqref{eq:Ftc} and \eqref{eq:Fsc} are independent of the scaling factor $f$, and we have $F_{\rm SC}/F_{\rm TC}=0.60$ for all five coils. 
Plots of the SCIF ratio versus inner diameter $2a_1$ of coils in Fig.~\ref{fig:Bsc_a1} show that the numerical data are roughly fitted by our theoretical results from Eq.~\eqref{eq:Bsc/Btc_alpha}, $|B_{\rm SC}/B_{\rm TC}| \propto 1/2a_1$. 
We notice in Fig.~\ref{fig:Bsc_a1} that our theoretical results tend to underestimate the SCIF ratio for large $2a_1$. 
Larger magnetic field is generated in larger coils, as we discussed earlier, and we speculate that the underestimation of the SCIF is due to the dependence of $J_{\rm c}(B)$ on the magnetic field strength $B$. 
The larger $J_{\rm c}(B)$ dependence leads to larger $\alpha_{\rm c}$ in Eq.~\eqref{eq:Bsc/Btc_alpha}, resulting in larger SCIF ratio.

\begin{figure}[tb]
 \center\includegraphics{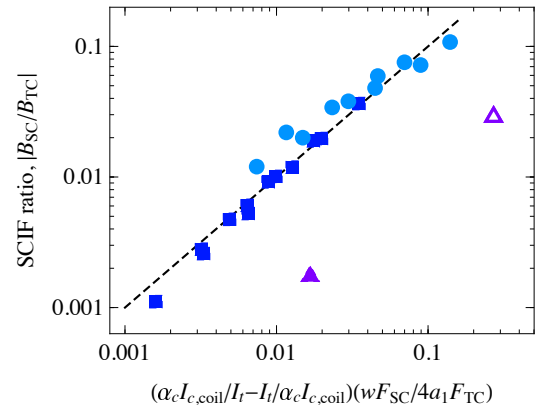}
\caption{%
SCIF ratio $|B_{\rm SC}/B_{\rm TC}|$ vs the theoretical SCIF ratio given by the right-hand side of Eq.~\eqref{eq:Bsc/Btc_alpha} with $\alpha_{\rm c}=1.5$. 
The symbols show the numerical results published by Ueda {\em et al}.~\cite{Ueda_15} (squares), Yanagisawa {\em et al}.~\cite{Yanagisawa_11} (circles), Amemiya and Akachi~\cite{Amemiya_08} (closed triangle), and Koyama {\em et al}.~\cite{Koyama_09} (open triangle). 
The dashed line corresponds to our theoretical results given by Eq.~\eqref{eq:Bsc/Btc_alpha} with $\alpha_{\rm c}=1.5$. 
}\label{fig:Bsc-theory_numerical}
\end{figure}

We examine the validity of our theoretical results by comparing them with published numerical results~\cite{Amemiya_08,Ueda_15,Yanagisawa_11,Koyama_09}. 
Figure~\ref{fig:Bsc-theory_numerical} shows the SCIF ratio $|B_{\rm SC}/B_{\rm TC}|$ versus the theoretical SCIF ratio given by the right-hand side of Eq.~\eqref{eq:Bsc/Btc_alpha}. 
The numerical data shown in Figs.~\ref{fig:Bsc_w} (square symbols) and \ref{fig:Bsc_a1} (circle symbols) are also plotted in Fig.~\ref{fig:Bsc-theory_numerical}, and those numerical data are roughly fitted by Eq.~\eqref{eq:Bsc/Btc_alpha} (dashed line). 
In contrast, the numerical data from Amemiya and Akachi~\cite{Amemiya_08} (closed triangle) and by Koyama {\em et al}.~\cite{Koyama_09} (open triangle) are much smaller than the theoretical values from Eq.~\eqref{eq:Bsc/Btc_alpha}. 
The ratio of the height to the outer diameter of the coils was $b/a_2=0.23$--$0.46$ for the coils investigated by Ueda {\em et al}.~\cite{Ueda_15}, $b/a_2=0.63$ for Yanagisawa {\em et al}.~\cite{Yanagisawa_11}, $b/a_2=1.65$ for Amemiya and Akachi~\cite{Amemiya_08}, and $b/a_2=1.62$ for Koyama {\em et al}.~\cite{Koyama_09}. 
The fully magnetized wire model neglecting the $z_k$ dependence of $|M_r|$ as in Eq.~\eqref{eq:Mr=Mmax} cannot be used for tall coils of $b/a_2>1$, whereas the model is acceptable for flat coils of $b/a_2<1$. 

The purpose of this paper is the rough evaluation of the SCIF rather than the precise calculation of the magnetic field in HTS coils. 
To this end, we adopted the fully magnetized model with constant critical current density given by Eqs.~\eqref{eq:Mr=Mmax}--\eqref{eq:Mmax-Ic,coil}. 
This model assumes oversimplified distribution of the magnetization, and the dependence of $J_{\rm c}$ on the local magnetic field is neglected, although the parameter $\alpha_{\rm c}$ implicitly reflects the field dependence of $J_{\rm c}$. 
This model, however, is sufficient for rough evaluation of the SCIF, and the resulting Eq.~\eqref{eq:Bsc/Btc_alpha} can well explain the dependence of the SCIF on the tape wire (Fig.~\ref{fig:Bsc_w}) and on the coil size (Fig.~\ref{fig:Bsc_a1}). 
Furthermore, the SCIF ratios in many numbers of coils with various dimensions are roughly fitted by our theoretical results, as shown in Fig.~\ref{fig:Bsc-theory_numerical}. 

In this paper, we theoretically investigated the SCIF in solenoid coils with superconducting tape wires. 
The direct relationship between the SCIF at the center of a coil and the distributed magnetization of tape wires is obtained as Eq.~\eqref{eq:Bsc}. 
A simple scaling law is derived: the SCIF ratio varies roughly in inverse proportion to the coil dimensions. 
Based on the fully magnetized wire model with the critical state model, the SCIF is simply given by Eqs.~\eqref{eq:Mmax-Ic,coil}--\eqref{eq:Bsc=+MmaxFsc}, and the SCIF ratio is then given by Eq.~\eqref{eq:Bsc/Btc_alpha} as functions of the tape width, the current load factor, and the coil dimensions. 
These theoretical results for SCIF agree roughly with the published results of the accurate numerical computation for flat coils of $b/a_2<1$ for which the fully-magnetized-tape model is valid, although our model overestimates the SCIF for tall coils of $b/a_2>1$. 
The present theoretical model is an oversimplification; however, Eq.~\eqref{eq:Bsc/Btc_alpha} is useful for the rough estimation of the SCIF, especially in the early stages of designing superconducting coils.

\acknowledgments 
We thank M. Furuse, K. Kajikawa, Y. Higashi, T. Wakuda, S. Yokoyama, M. Breschi, and S. Awaji for stimulating discussions. 
This paper is based on results obtained from a project commissioned by the New Energy and Industrial Technology Development Organization (NEDO).


\begin{thebibliography}{99}
\bibitem{Trociewitz_11}
U. P. Trociewitz, M. D. Canassy, M. Hannion, D. K. Hilton, J. Jaroszynski, P. Noyes, Y. Viouchkov, H. W. Weijers, and D. C. Larbalestier, 
Appl. Phys. Lett. {\bf 99}, 202506 (2011).
\bibitem{Maeda_13}
H. Maeda and Y. Yanagisawa, 
IEEE Trans. Appl. Supercond. {\bf 24}, 4602412 (2014).

\bibitem{Miyazaki_16}
H. Miyazaki, S. Iwai, Y. Otani, M. Takahashi, T. Tosaka, K. Tasaki, S. Nomura, T. Kurusu, H. Ueda, S. Noguchi, A. Ishiyama, S. Urayama, and H. Fukuyama, 
Supercond. Sci. Technol. {\bf 29}, 104001 (2016).
\bibitem{Yokoyama_17}
S. Yokoyama, J. Lee, T. Imura, T. Matsuda, R. Eguchi, T. Inoue, T. Nagahiro, H. Tanabe, S. Sato, A. Daikoku, T. Nakamura, Y. Shirai, D. Miyagi, and M. Tsuda, 
IEEE Trans. Appl. Supercond. {\bf 27}, 4400604 (2017).

\bibitem{Yanagisawa_14}
Y. Yanagisawa, R. Piao, S. Iguchi, H. Nakagome, T. Takao, K. Kominato, M. Hamada, S. Matsumoto, H. Suematsu, X. Jin, M. Takahashi, T. Yamazaki, and H. Maeda, 
J. Magn. Reson. {\bf 249}, 38 (2014).
\bibitem{Nishijima_16}
G. Nishijima, S. Matsumoto, K. Hashi, S. Ohki, A. Goto, T. Noguchi, S. Iguchi, Y. Yanagisawa, M. Takahashi, H. Maeda, T. Miki, K. Saito, R. Tanaka, and T. Shimizu, 
IEEE Trans. Appl. Supercond. {\bf 26}, 4303007 (2016).

\bibitem{Amemiya_12}
N. Amemiya, K. Takahashi, H. Otake, T. Nakamura, Y. Mori, T. Ogitsu, K. Koyanagi, A. Osanai, T. Yoshiyuki, K. Noda, and M. Yoshimoto, 
Physica C {\bf 482}, 74 (2012).
\bibitem{Ueda_13}
H. Ueda, M. Fukuda, K. Hatanaka, T. Wang, A. Ishiyama, and S. Noguchi, 
IEEE Trans. Appl. Supercond. {\bf 23}, 4100805 (2013).
\bibitem{Rossi_15}
L. Rossi, A. Badel, M. Bajko, A. Ballarino, L. Bottura, M.M.J. Dhalle, M. Durante, P. Fazilleau, J. Fleiter, W. Goldacker, E. Haro, A. Kario, G. Kirby, C. Lorin, J. van Nugteren, G. de Rijk, T. Salmi, C. Senatore, A. Stenvall, P. Tixador, A. Usoskin, G. Volpini, Y. Yang, and N. Zangenberg, IEEE Trans. Appl. Supercond. {\bf 25}, 4001007 (2015).

\bibitem{Benkel_17}
T. Benkel, Y. Miyoshi, X. Chaud, A. Badel, and P. Tixador, 
Eur. Phys. J. Appl. Phys. {\bf 79}, 30601 (2017).

\bibitem{Hemmi_07}
T. Hemmi, N. Yanagi, K. Seo, G. Bansal, K. Takahata, and T. Mito, 
IEEE Trans. Appl. Supercond. {\bf 17}, 2422 (2007).
\bibitem{Amemiya_08}
N. Amemiya and K. Akachi, 
Supercond. Sci. Technol. {\bf 21}, 095001 (2008).
\bibitem{Ahn_09}
M. C. Ahn, T. Yagai, S. Hahn, R. Ando, J. Bascunan, Y. Iwasa, 
IEEE Trans. Appl. Supercond. {\bf 19}, 2269 (2009).
\bibitem{Koyama_09}
Y. Koyama, T. Takao, Y. Yanagisawa, H. Nakagome, M. Hamada, T. Kiyoshi, M. Takahashi, and H. Maeda, 
Physica C {\bf 469}, 694 (2009).

\bibitem{Yanagisawa_09}
Y. Yanagisawa, H. Nakagome, Y. Koyama, R. Hu, T. Takao, M. Hamada, T. Kiyoshi, M. Takahashi and H. Maeda, 
Physica C {\bf 469}, 1996 (2009).
\bibitem{Kajikawa_11}
K. Kajikawa and K. Funaki, 
Supercond. Sci. Technol. {\bf 24}, 125005 (2011).
\bibitem{Yanagisawa_12}
Y. Yanagisawa, Y. Kominato, H. Nakagome, T. Fukuda, T. Takematsu, T. Takao, M. Takahashi, and H. Maeda, 
AIP Conf. Proc. {\bf 1434}, 1373 (2012).
\bibitem{Yanagisawa_15}
Y. Yanagisawa, Y. Xu, X. Jin, H. Nakagome, and H. Maeda, 
IEEE Trans. Appl. Supercond. {\bf 25}, 6603705 (2015).


\bibitem{Ueda_15}
H. Ueda, J. Saito, Y. Ariya, A. Mochida, T. Wang, X. Wang, K. Agatsuma, and A. Ishiyama, 
IEEE Trans. Appl. Supercond. {\bf 25}, 6603205 (2015).

\bibitem{Pardo_16}
E. Pardo, Supercond. Sci. Technol. {\bf 29}, 085004 (2016).

\bibitem{Jackson_75}
J. D. Jackson {\em Classical Electrodynamics}, 2nd ed. (Wiley, New York, 1975).

\bibitem{Bean_62}
C. P. Bean, Phys. Rev. Lett. {\bf 8} 250 (1962).
\bibitem{Bean_64}
C. P. Bean, Rev. Mod. Phys. {\bf 36} 31 (1964).

\bibitem{Kajikawa_15}
K. Kajikawa, G. V. Gettliffe, Y. Chu, D. Miyagi, T. P. Lecrevisse, S. Hahn, J. Bascunan, and Y. Iwasa, 
IEEE Trans. Appl. Supercond. {\bf 25}, 4300305 (2015).

\bibitem{Wilson83}
M. N. Willson, {\em Superconducting Magnets} (Oxford: Clarendon, 1983).

\bibitem{Yanagisawa_11}
Y. Yanagisawa, Y. Kominato, H. Nakagome, R. Hu, T. Takematsu, T. Takao, D. Uglietti, T. Kiyoshi, M. Takahashi, and H. Maeda, 
IEEE Trans. Appl. Supercond. {\bf 21}, 1640 (2011).

\end{thebibliography}
\end{document}